\documentstyle[12pt]{article}

\columnwidth\textwidth

\begin{document}

\newcommand{\be}{\begin{equation}}
\newcommand{\ee}{\end{equation}}
\newcommand{\beann}{\begin{eqnarray*}}
\newcommand{\eeann}{\end{eqnarray*}}
\newcommand{\bea}{\begin{eqnarray}}
\newcommand{\eea}{\end{eqnarray}}
\newcommand{\nn}{\nonumber}
\newcommand{\ben}{\begin{enumerate}}
\newcommand{\een}{\end{enumerate}}
\newtheorem{df}{Definition}
\newtheorem{thm}{Theorem}
\newtheorem{lem}{Lemma}
\newtheorem{prop}{Proposition}
\begin{titlepage}

\noindent
\hspace*{11cm} BUTP-95/36 \\
\vspace*{1cm}
\begin{center}
{\LARGE Time evolution and observables in constrained systems}

\vspace{2cm}

P. H\'{a}j\'{\i}\v{c}ek \\
Institute for Theoretical Physics \\
University of Bern \\
Sidlerstrasse 5, CH-3012 Bern, Switzerland
\\ \vspace*{2cm}

December 1995 \\ \vspace*{1cm}

\nopagebreak[4]

\begin{abstract}
The discussion is limited to  first-class parametrized systems, where the
definition of time evolution and observables is not trivial, and to finite
dimensional systems in order that technicalities do not obscure the conceptual
framework. The existence of reasonable true, or physical, degrees of freedom is
rigorously defined and called {\em local reducibility}. A proof is given that
any locally reducible system admits a complete set of perennials. For locally
reducible systems, the most general construction of time evolution in the
Schroedinger and Heisenberg form that uses only geometry of the phase space
is described. The time shifts are not required to be
symmetries. A relation between perennials and observables of the Schroedinger
or Heisenberg type results: such observables can be identified with certain
classes of perennials and the structure of the classes depends on the time
evolution. The time evolution between two non-global transversal surfaces is
studied. The problem is posed and solved within the framework of the ordinary
quantum mechanics. The resulting non-unitarity is different from that known in
the field theory (Hawking effect): state norms need not be preserved so that
the system can be lost during the evolution of this kind.

\end{abstract}

\end{center}

\end{titlepage}

\section{Introduction}
A truly intriguing feature of the general relativity is the lack of any fixed
background spacetime that would serve as a stage for its dynamics. There are
many different spacetimes that solve Einstein's equations; but the time
evolution of the given gravitating system in the strict sense that we are used
to from the study of other systems does not take place in any of them. The
discussion of these problems is somewhat confined to the group of people who
are trying to quantize the theory and the issue is called ``the problem of
time in quantum gravity'' (see, e.\ g.\ \cite{isham-rev2} and
\cite{kuch-prehled}). However, even within the classical version of Einstein's
theory, the concept of time evolution and the related one of observable which
would be sufficiently closely analogous to that of other models of theoretical
physics are either not suitable for Einstein's theory or not yet completely
developed.

	An impressive work in this direction has been done by Kucha\v{r}. His
method is to reconstruct the naked spacetime manifold (that is, without
metric) from the phase space by separating the kinematical variables from the
dynamical ones; the kinematical variables describe the position at the naked
manifold and the dynamical variables become observables which evolve along it.
The approach of the
present paper---the so-called perennial formalism---owes much to Kucha\v{r}'
ideas. However, it attempts to construct the dynamics directly within the
phase space so that no form of spacetime is needed at any stage. The
construction is based on some well-known (even
very old) ideas. First of them is Dirac's theory of the so-called ``three
forms of relativistic dynamics'' \cite{dirac} for a system of massive
particles in Minkowski spacetime. This is based on one hand on the
Poincar\'{e} group or algebra and on the other hand on three kinds of surfaces
defining the three forms.  Although Dirac considered these surfaces as lying
in the spacetime, each of them defines a unique surface in the phase space and
the properties of these surfaces that are essential for the method to work can
even more easily be understood within the phase space: they are ``transversal
surfaces'' (see \cite{honeff}); any reasonable system possesses such surfaces.
Similarly the Poincar\'{e} group or algebra is a structure which can be found
in the phase space of reasonable systems: it is a group of symmetries or an
algebra of perennials. These are the basic notions used by some
``modern'' methods of quantizing the parametrized systems, in particular the
group and algebraic quantization (\cite{isham}, \cite{rov-group} and
\cite{bluebook}). The corresponding generalization of Dirac's idea to any
finite-dimensional parametrized system has been given in \cite{timelevels}; an
infinite-dimensional system (the massive scalar field in curved background
spacetime) was studied in \cite{H+I}, where the geometric theory of
infinite-dimensional Hamiltonian systems by Marsden and his collaborators
(\cite{C+M} and \cite{F+M}) helped to solve the problems \cite{I+H}.

	Roughly, the present paper contains three new ideas. First, a
distinction between integrals
of motion and perennials is recognized; this yields several useful
insights. Second, a general construction of time evolution for
parametrized systems is given and the importance of the time evolution for the
notion of observable is clarified.  Third, the time evolution between two
non-global transversal surfaces is considered as an exercise in, and is solved
using just the tools of, ordinary quantum mechanics. The plan of the paper
is as follows.

	Section \ref{sec:singular} contains an extension of the notion of
perennial and symmetry to the so-called singular perennial and symmetry, which
is necessary for the method to work for non-global transversal surfaces of
certain kind that are often met (for example inextensible transversal surfaces
which are not global). The property of local reducibility, that is the
existence of reasonable true degrees of freedom, is rigorously defined and
shown to imply the existence of a complete system of perennials. A relation
between integrals of motion of (unconstrained) Hamiltonian systems and
perennials of the corresponding parametrized systems is clarified. In
particular, ``chaotic'' Hamiltonian systems that do not admit any integral of
motion except for the Hamiltonian, do admit a complete system of perennials if
parametrized.  Section \ref{sec:quant} recalls briefly the quantization theory
as given in \cite{timelevels}, and it brings some improvements, especially the
use of universal enveloping algebras. Section \ref{sec:evol} contains a
construction of time evolution using the so-called time shifts, which
is, in a sense explained in subsection \ref{sec:general}, the most general
time evolution possible. In particular, no symmetry is now neccessary for such
a construction. This generality should not be understood in the sense,
however, that each possible time evolution which can be constructed for a
given system according to our prescription is sensible---a choice has to be
done.  A differentiable one-dimensional case of time evolution is studied in
subsection \ref{sec:cont}, where the general form of the Heisenberg and
Schroedinger equations of motion is derived. An example shows that our new
construction contains time evolution that can also be obtained by the method
of reduction (see e.\ g.\ \cite{I+B} or \cite{moncrief}). In the final
subsection \ref{sec:nonglob} of the section
\ref{sec:evol}, we investigate the evolution between two non-global transversal
surfaces. In fact, an example of such an evolution for a field system was
studied in (\cite{H+I}): the Hawking effect. It was shown that a careful
consideration of domains and ranges of time shifts can explain the well-known
non-unitarity of time evolution in this case. However, in the field system
case, the evolution just looses information; the normalization of states is
preserved, because (roughly speaking) even the state of no excitation is a
normalized state (vacuum) of the system.
Surprisingly, the situation is worse for finite dimensional
systems, where the non-arrival of the system at a final transversal surface
must be interpreted as a loss of the system during the evolution---the
non-unitarity is then  of a different type (not preserving norms).
However, the conclusion about the
non-unitarity of the evolution follows necessarily once the choice of the two
transversal surfaces is met. In section \ref{sec:peren}, we discuss the notion
of observable and its relation to that of perennial. It turns out that this
notion is related to, but not completely identical with, that of
``evolving constant of motion'' by Carlo Rovelli \cite{rov+haj}. Thus, the
observables are not identical with perennials (in general); formally, they are
classes of perennials. These observables are of the ordinary quantum
mechanical type; they are measured each by a distinguished measurement
process or apparatus that is well-defined independently of time (``the same
measurement at different times'').
Finally, in section \ref{sec:patch}, we illustrate the construction of the time
evolution between two non-global transversal surfaces using a simple model of
a completely solvable system that does not admit global transversal surfaces.

\section{Singular perennials and symmetries}
\label{sec:singular}
In this section we generalize the notion of perennial and of symmetry in a way
that will lead to simplifications in our subsequent work on non-global
transversal surfaces.

	Let us first recall the few basic facts about the first-class
parametrized systems (for details see \cite{timelevels}). We restrict
ourselves to finite-dimensional models so the phase space will be a
\(2N\)-dimensional manifold \(\tilde{\Gamma}\) with a symplectic form
\(\tilde{\Omega}\). The dynamics is determined by the constraint surface
\(\Gamma\) of a special kind (for a first-class system): \(\Gamma\) is a
\((2n-\nu)\)-dimensional submanifold of \(\tilde{\Gamma}\) such that the
pull-back \(\Omega\) of \(\tilde{\Omega}\) to \(\Gamma\) is a pre-symplectic
form whose singular subspace \(L_p\) at the point \(p\in\Gamma\) has
the dimension \(\nu\) for all \(p\). Then \(L_p\) is an integrable
distribution on \(\Gamma\); the maximal integral manifolds \(\gamma\)
of \(L_p\) are called c-orbits. Each c-orbit represents a unique maximal
classical solution in all possible gauges and foliations. Each point
\(p\in\Gamma\) lies at exactly one c-orbit, which will be denoted by
\(\gamma_p\).

	A perennial is defined as a differentiable function
\(o:\tilde{\Gamma}\mapsto\mbox{\bf R}\) that is constant along each c-orbit.
Our generalization will allow perennials to be \(C^{\infty}\) only on a
subset,
\({\cal D}(o)\) of \(\tilde{\Gamma}\), the so-called {\em domain} of
\(o\). The set
\({\cal D}(o)\) must have the following properties:
\begin{enumerate}
\item \({\cal D}(o)\) is open in \(\tilde{\Gamma}\),
\item \(\Gamma\subset\overline{{\cal D}(o)}\).
\end{enumerate}
Such perennials will be called {\em singular}.
Let \(o_1\) and \(o_2\) be two perennials with domains \({\cal D}(o_1 )\) and
\({\cal D}(o_2 )\). Then the linear combination, functional multiplication and
Poisson brackets of \(o_1\) and \(o_2\) are all well-defined on  \({\cal D}(o_1
)\cap{\cal D}(o_2 )\). As this set has again the properties of a domain, the
three operations will result in singular perennials. All singular
perennials form a Poisson algebra which we denote by \({\cal P}\).

	A useful objects will be the projectors associated with some open
subsets
of \(\tilde{\Gamma}\); we define them as maps in \({\cal P}\). Let \({\cal
D}\) be an
open subset of \(\tilde{\Gamma}\) with the property: if \(p\in({\cal
D}\cap\Gamma)\), then \(\gamma_p\in{\cal D}\). Let \(\chi_{\cal
D}:\tilde{\Gamma}\mapsto\mbox{\bf R}\) be the characteristic function of
\({\cal D}\), that is
\beann
	\chi_{\cal D}(p) & = & 1\quad \forall p\in{\cal D}, \\
	\chi_{\cal D}(p) & = & 0\quad \forall
	p\in\tilde{\Gamma}\setminus\bar{\cal D}.
\eeann
Then \(\chi_{\cal D}\) is a (singular) perennial. \(\chi_{\cal D}\) defines a
map \(\Pi_{\cal D}:{\cal P}\mapsto{\cal P}\) by \(\Pi_{\cal D}(o)=\chi_{\cal
D}o\) for all \(o\in{\cal P}\). It follows easily that \({\cal D}(\Pi_{\cal
D}(o))=({\cal D}\cup(\tilde{\Gamma}\setminus\bar{\cal D}))\cap{\cal D}(o)\),
that
\(\Pi_{\cal D}({\cal P})={\cal P}_{\cal D}\) is a Poisson algebra, and that
\(\Pi_{\cal D}\) is a Poisson algebra homomorphism. \(\Pi_{\cal D}\) has all
properties of a projection operator.

	The next notion that plays an important role in the perennial
formalism is that of transversal surface. Recall that such a surface is a
submanifold \(\Gamma_1\) of the constraint manifold \(\Gamma\) which has no
common tangent vectors with the c-orbits (except for zero vector) and which
intersects each c-orbit in at most one point.
The set \({\cal D}(\Gamma_1 ):= \{p\in\Gamma|\gamma_p\cap\Gamma_1 \neq
\emptyset\}\) is called domain of \(\Gamma_1\) and
\(\Gamma_1\) is called a global trasversal surface, if \({\cal
D}(\Gamma_1 )=\Gamma\). The pull-back \(\Omega_1\) of the symplectic form
\(\Omega\) to \(\Gamma_1\) is non-degenerate so that the pair \((\Gamma_1
,\Omega_1)\) is a
symplectic manifold; we denote the corresponding Poisson brackets by \(\{\cdot
,\cdot\}_1\). This symplectic manifold can be considered as the phase space of
the corresponding {\em reduced} system; in particular, the number of {\em true
degrees of freedom} is half the dimension of \(\Gamma_1\).
Symmetries and perennials can be projected to transversal
surfaces: Let \(i_1\) be the embedding of \(\Gamma_1\) in \(\tilde{\Gamma}\)
and \(\pi_1 : \Gamma\mapsto\Gamma_1\) be defined by
\(\pi_1 (p)=\gamma_p\cap\Gamma_1\); \(\pi_1\) is called {\em projector
associated with} \(\Gamma_1\). Then each symmetry \(\varphi\) which
preserves the domain of \(\Gamma_1\) defines a map \(a_1 (\varphi
):\Gamma_1\mapsto\Gamma_1\) by \(a_1 (\varphi )(p)=\pi_1 (\varphi (p))\). The
map \(a_1\) preserves the composition of the symmetries; thus it defines an
action of groups of symmetries provided that all elements of the group
preserve the domain
of \(\Gamma_1\). If \(o\) is a perennial, then \(o_1 = i_1^* o\) is a function
on \(\Gamma_1\); \(i_1^*\) preserves the linear combination, product of
functions and the Poisson bracket, i.\ e.\ \(i_1^*\{o,o'\} = \{i_1^* o,i_1^*
o'\}_1\). Thus, \(i_1^*\) is a homomorphism of Poisson algebras. For details
see \cite{timelevels}.

	The definition of the first-class parametrited systems as given above
and in \cite{timelevels} is too general for physicist's purposes. Generically,
such a system cannot be reduced even locally, that is, there will be no
transversal surfaces in any neighbourhood of any point of \(\Gamma\). To
exclude this pathology, we restrict ourselves to the {\em locally reducible}
systems, which can be defined as follows.
\begin{df}
A first-class parametrized system \((\tilde{\Gamma},\tilde{\Omega},\Gamma )\)
is called locally reducible, if a dense open subset of \(\Gamma /\gamma\) is a
quotient manifold (not necessarily Hausdorff).
\end{df}
For the definition of quotient manifolds see, e.\ g.\ \cite{B+C}. In
particular, the natural projection \(\pi : \Gamma\mapsto \Gamma /\gamma\) is a
submersion. Then, as shown in \cite{B+C}, there is a differentiable section of
\(\pi\) through any point of \(\Gamma\). A section of \(\pi\) is a
map \(\psi : \Gamma /\gamma\mapsto \Gamma\) such that
\(\pi\circ\psi\) is the identity on the domain of \(\psi\) (which is
necessarily Hausdorff). This implies easily that the image of \(\psi\) is a
transversal surface. Inversely, suppose that every point \(p \in \Gamma\) lies
at some transversal surface and that the associated projectors are
differentiable. Then the quotient set can
be given a quotient manifold structure by pasting all these transversal
surfaces by their assosiated projectors in the overlapping domains \({\cal
D}(\Gamma_i)\cap {\cal D}(\Gamma_j)\). This justifies the definition.
The general relativity may be locally reducible (see \cite{F+M}).

	The locally reducible systems have the nice property that they admit
complete systems of (singular) perennials. We will say that a system of
perennials is complete, if it separates separable c-orbits; the c-orbits
\(\gamma_1\) and \(\gamma_2\) are separable if there is a continuous perennial
\(o\) such that \(o(\gamma_1)\neq o(\gamma_2 )\). Indeed, in the special case
that \(\Gamma /\gamma\) is Hausdorff, we can construct such a system as
follows. According to the classical theorem by Whitney, \(\Gamma /\gamma\) can
be globally embedded in \(\mbox{\bf R}^{\kappa}\), where \(\kappa = 4N - 4\nu
+ 1\) (because the dimension of \(\Gamma /\gamma\) is \(2N - 2\nu\)
\cite{sternberg}). Let \(X^k\), \(k = 1,\ldots ,\kappa\) be the natural
coordinates on \(\mbox{\bf R}^{\kappa}\) and let \(\Phi :\Gamma /\gamma\mapsto
\mbox{\bf R}^{\kappa}\) be the embedding. Then \(X^k\circ\Phi\circ\pi\), \(k =
1, \ldots ,\kappa\) is a complete system of perennials on
\(\Gamma\). Moreover, the gradients of all elements of the system span the
subspace of \(T^*_p\Gamma\) that is transversal to
\(T_p\gamma\) at each point \(p\in\Gamma\). In the general case, when \(\Gamma
/\gamma\) need not be Hausdorff, one can find a complete system of singular
perennials as follows. Let us recall that any non-Hausdorff manifold \(M\) can
be decomposed in its maximal Hausdorff submanifolds \(M_i\): any point of \(M\)
lies at some Hausdorff submanifold of \(M\) (namely, the corresponding chart),
and all Hausdorff submanifolds of \(M\) form a partially ordered set with the
right properties so that one easily obtains the desired existence. Let, then,
\(\Gamma
/\gamma = \bigcup_i M_i\) be this decomposition of \(\Gamma /\gamma\). For
each \(M_i\), a complete set of perennials can be constructed according to the
procedure desribed above. The functions we find in this way, however, need not
possess differentiable extensions to the boundaries \(\partial M_i\) of
\(M_i\) in \(\Gamma /\gamma\) (for examples, see section \ref{sec:patch} and
\cite{patchI}). This motivates our introduction of singular perennials: we can
define such perennials everywhere on \(\Gamma /\gamma\) by setting them equal
to zero in \(\Gamma /\gamma\setminus\overline{\pi^{-1}M_i}\). Working this out
for each \(i\), one obtains a (hopefully finite) complete set of singular
perennials.

	The construction of perennials in \({\cal D}(\Gamma_1)\), where
\(\Gamma_1\) is a transversal surface can start from \(\Gamma_1\) instead of
\(\Gamma /\gamma\). Indeed, \(\pi|_{\Gamma_1}\) is a diffeomorphism between
\(\Gamma_1\) and \(\pi\Gamma_1\). Thus, a differentiable function \(o_1\) on
\(\Gamma_1\) can be pulled back by \((\pi|_{\Gamma_1})^{-1}\) to
\(\pi\Gamma_1\) and the resulting perennial \(o\) is given by \(o =
o_1\circ(\pi|_{\Gamma_1})^{-1}\circ\pi = o_1\circ\pi_1\), because \(\pi_1 =
(\pi|{\Gamma_1})^{-1}\circ\pi\). \(o\) will be referred to as {\em defined by
the initial datum} \(o_1\) {\em at} \(\Gamma_1\).

	To prevent misunderstanding, some comment is in order. On one hand,
perennials can be considered as ``integrals of motion'' of the system. On the
other, many completely regular and physically reasonable Hamiltonian systems
do not admit any integrals of motion. This seems to be a paradox. In order to
remove the paradox,
we must become a little more precise. A Hamiltonian system
\((V,\Omega ,H)\) consists of a symplectic manifold \((V,\Omega )\) and a
differentiable function \(H\) whose Hamiltonian vector field on \(V\) is
complete. An integral of motion is a function on \(V\) which is constant along
the orbits of \(H\). It has been shown in \cite{robinson} that such systems
{\em generically} do not admit any integral of motion independent from the
Hamiltonian. For example, the movement of a
material point on a frictionless surface \(\Sigma\) without external forces is
such a system, if
\(\Sigma\) is a compact Riemannian manifold with constant negative curvature
(\(V =
T^*\Sigma\), see \cite{arnold}); there is nothing pathological with this
system. \((V, \Omega ,H)\) is {\em no} constrained system, however. To obtain
a first-class
parametrized system from it that will describe the same motion, one must
{\em parametrize} it. This is the following procedure. Let \(\tilde{V}:=
V\times \mbox{\bf R}^2\) and let the natural coordinates on \(\mbox{\bf R}^2\)
be \(t\) and \(p_t\). Define the symplectic form \(\tilde{\Omega}\) on
\(\tilde{V}\) by \(\tilde{\Omega} := \Omega + dp_t\wedge dt\) and the
constraint surface \(\Gamma\) by the equation \(p_t + H = 0\). One easily
verifies that the
corresponding c-orbits, if projected down to \(V\) by the natural projection
in the cartesian product \(V\times\mbox{\bf R}^2\), coincide with the dynamical
trajectories of \((V,\Omega ,H)\). However, this correspondence is
many-to-one; c-orbits that are mapped on the same trajectory are
obtained from different time parametrizations of the trajectory.
Thus, perennials of \((\tilde{V}, \tilde{\Omega},
\Gamma )\) need not coincide with the integrals of motion of \((V,\Omega
,H)\): an integral defines a perennial, but a perennial need not determine any
integral. Let us show
that the system \((\tilde{V},\tilde{\Omega},\Gamma )\) is locally
reducible. For this aim, we define the map \(\Psi : (V\times\mbox{\bf
R})\mapsto \Gamma\) by \(\Psi(p,t) = (\Phi_t(p), t, H(p))\), where \(\Phi_t\)
is the flow of the Hamiltonian
vector field of \(H\) on \(V\). \(\Psi\) is a diffeomorphism, because \(\Phi_t
: V\mapsto V\) is a diffeomorphism for each \(t\in \mbox{\bf R}\) and
\(\Phi_t(p) : \mbox{\bf R} \mapsto V\) is a differentiable curve
at each \(t\in \mbox{\bf R}\) and for each \(p\in V\). Moreover,
\(\Psi(p,\mbox{\bf
R})\) is the c-orbit through the point \((p, 0, H(p))\) of the surface \(t =
0\) in \(\Gamma\) for any \(p\in V\). Consider the map \(\pi_V \circ
\Psi^{-1}\), where \(\pi_V : (V \times \mbox{\bf R}) \mapsto V\) is the
natural projection of a Cartesian product of manifolds. \(\pi_V \circ
\Psi^{-1}\) maps all points of any c-orbit to just one point of \(V\). Thus,
\(\pi_V \circ \Psi^{-1}\) can be considered as mapping \(\Gamma /\gamma\) to
\(V\); as such it is a bijection. We may use \(\pi_V \circ
\Psi^{-1}\) to define a manifold structure on \(\Gamma /\gamma\); with this
structure, \(\Gamma /\gamma\) is a quotient manifold. Indeed, \(\pi :
\Gamma \mapsto \Gamma /\gamma\) can be identified with \(\pi_V \circ
\Psi^{-1}\), and this is a submersion. As a byproduct, we have that \(\Gamma_0
:= \{(p, 0, H(p))|p \in V\}\) is a global transversal surface.

	To summarize: this example shows that parametrizing a Hamiltonian
system always results in a constrained system with a complete set of
perennials independently of how many integrals of motion the Hamiltonian
system possesses. Clearly, a parametrized system without a complete system of
perennials has a different status than a Hamiltonian system without integrals
of motion: the former is pathological, the latter is not. The locally
reducible systems are, however, rather rare among all first-class parametrized
systems. To understand that, the following observation is useful. Formally,
another parametrized system can be constructed from the Hamiltonian system
\((V, \Omega ,H)\): this is
\((V, \Omega ,\Gamma')\), where \(\Gamma'\) is defined by the equation \(H =
E\) and \(E \in H(V)\). Such a parametrized system is not locally reducible,
if \((V, \Omega ,H)\) does not admit a complete system of integrals of motion
(that is, separating dynamical trajectories).

\section{Quantization}
\label{sec:quant}
In this section, we wish to combine the algebraic Ashtekar method of
quantization with the group method by Isham and simultaneously allow for the
singular perennials.

	Let \({\tilde G}_0\) be a Lie group of symmetries; that is, each
element of \({\tilde G}_0\) is a symmetry, and there is a common invariant
domain, \({\cal D}({\tilde G}_0 )\) of all elements of \({\tilde G}_0\)
such that \(\overline{{\cal D}(\tilde{G}_0 )\cap\Gamma}=\Gamma\).
Recall that a group \(G\) is called almost transitive if there is a
c-orbit \(\gamma\) such that \(\overline{G(\gamma )} = \Gamma\).
All elements of \({\tilde G}_0\) that leave the c-orbits invariant form a
normal
subgroup \(N\). Let \({\tilde S}_0\) be the Lie algebra of \({\tilde G}_0\).
The action of \({\tilde G}_0\) on \({\tilde{\Gamma}}\) enables us to realize
\({\tilde S}_0\) as a Lie algebra of vector fields on \({\cal D}({\tilde G}_0
)\). Let us call the group \({\tilde G}_0\) {\em Hamiltonian}, if all these
vector fields are globally Hamiltonian. Then each element of \({\tilde S}_0\)
determines a unique class \(\{o\}\) of perennials (each two elements of the
class differ by a constant function). These perennials will in general be
singular, but they will have a common domain containing \({\cal D}({\tilde
G}_0 )\). One can either choose representatives of the classes \(\{o\}\) in
such a way that they form a Lie algebra \({\tilde S}\) with respect to the
Poisson bracket---and which is then isomorph to the algebra \({\tilde
S}_0\)--- or, if this is not possible, that they generate the Lie algebra
\({\tilde S}\) which is isomorph to a central extension of \({\tilde
S}_0\). Let \({\tilde G}\) be the Lie group which is obtained from \({\tilde
G}_0\) by the corresponding central extension; then \({\tilde G}\) has a
well-defined action on \(\Gamma\), given by that of \({\tilde G}_0\) and by
the requirement that the central elements act trivially. One can show
(\cite{timelevels}) that \(N\) is still a normal subgroup of \({\tilde G}\).
Thus if we assume that
\begin{description}
\item[(a)] \({\tilde G}_0\) is almost transitive,
\item[(b)] \({\tilde G}_0\) is Hamiltonian,
\item[(c)] \(N\) is a closed subgroup,
\end{description}
then \(G:={\tilde G}/N\) is a Lie group; we call \(G\) {\em first-class
canonical
group} (FCC group). FCC subgroup of \(G\) is a subgroup which itself satisfies
the conditions a, b, and c above. The quantum theory is to be constructed
via some representations of the FCC group.

	The Lie group \(N\) determines the Lie algebra \(I_S\) of perennials;
\(I_S\) is a Lie ideal of \({\tilde S}\) and it consists of
all elements of \({\tilde S}\) which vanish at \(\Gamma\). Then \(S:={\tilde
S}/I_S\) is a Lie algebra. If we replace the point {\bf (a)} of the definition
of FCC group by
\begin{quote}
	{\bf (a')} \(S\) is a complete system of perennials,
\end{quote}
then \(S\) is called {\em the algebra of elementary perennials}. This algebra
will satisfy (cf. \cite{timelevels}) the following requirements
\begin{description}
\item[(c)] \(S\) is a Lie algebra with respect to the operations of linear
combination and Poisson bracket (these operations are well-defined for the
classes of perennials in \(S\));
\item[(d)] \(S\) is a complete system of perennials;
\item[(e)] let \({\cal D}(S)=\bigcap_{o\in\tilde{S}}{\cal D}(o)\) and let
\(\xi_o\) be the
Hamiltonian vector field of the function \(o\); then \(\xi_o\) is complete in
\({\cal D}(S)\) for all \(o\in S\). \({\cal D}(S)\) is called the {\em common
invariant domain} of \(S\).
\end{description}
Clearly, \({\cal D}(S)\) coincides with \({\cal D}({\tilde G}_0 )\). An
important observation is that each element of \(S\)---which is a class of
perennials---defines exactly one function on \(\Gamma\) (which is constant
along c-orbits). Another observation is that a complete system of perennials
(whose existence has been shown in section \ref{sec:singular}) does not
necessarily form an algebra of elementary perennials: the Hamiltonian vector
fields need not be complete, and the algebra need not close. There are
symplectic manifolds that do not admit any {\em finite} system of functions
that separate points, whose elements possess complete Hamiltonian vector
fields, and whose Poisson-bracket algebra closes. An example is an
orientable two-dimensional Riemannian manifold of genus two (sphere with two
handles), the symplectic form being the volume form. Still, there is a finite
set of functions that separates points of this manifold.

	The last key object of the the classical part of the theory is the
universal enveloping algebra \(A\) of \(S\). This algebra \(A\) is a
counterpart of the `abstract associative algebra' introduced by Ashtekar
(\cite{bluebook}). \(A\) is needed for a formulation of some important
conditions on the representations of the FCC group. These conditions---the
so-called {\em relations}---come about because the elements of \(S\)
considered as functions on \(\Gamma\) often are functionally dependent; it
holds e.\ g.\ that \(F(o_1 ,\ldots ,o_k ) = 0\) for \(o_1\in
S,\ldots ,o_k\in S\). We would like to transfer these relations
into the quantum theory. The popular way to do that is to identify \(F\) with
an
element of the algebra \(A\). This will be possible if \(F\) is a polynomial.
Even if \(F\) is a real analytic function, one can define \(F\) by a series;
one can extend the algebra \(A\) by formal series' to an associative algebra
\(\bar{A}\) (cf. \cite{jacobson}) and then try to place the series for \(F\) in
\(\bar{A}\). However, each such identification is a particular choice of factor
ordering, so one has to solve the `factor ordering problem' in each case
(there are always some reasonable requirements on the physical factor
ordering, see e.\ g.\ \cite{H+K}, but the factor ordering is still not
uniquely determined in many cases, must be chosen and represents another
ambiguity in the way from a classical to the quantum theory). Suppose that
this problem is solved. Then we have some elements of the algebra
\(\bar{A}\)---which will again be called relations---that should be
represented by zero operators. It can happen that some of the relations lies
in the center of \(\bar{A}\); this was observed by Pohlmayer in the cases of a
massive relativistic particle on Minkowski spacetime and of the string theory
\cite{pohlmayer}.  In this form, some constraints may reappear in the quantum
theory.

	The last step in the quantization is to find a unitary representation
\(R\) of the Lie group \(G\) on a Hilbert space \(K\) that satisfies the
conditions
\begin{enumerate}
\item the representation \(R\) of all FCC subgroups of \(G\) is irreducible;
\item all relations are represented by zero operators.
\end{enumerate}
The second conditions is sensible, because any unitary representation of a Lie
group will induce a representation of its Lie algebra by operators which have
a common linear invariant domain in the representation space; this domain
is the well-known G\aa rding subspace. Thus, the representation of the Lie
algebra can be extended to that of the universal enveloping algebra. In
addition, the operators representing the elements of the Lie algebra are
essentially self-adjoint on the G\aa rding domain (representations
of topological groups are automatically assumed to be continuous, cf.\
\cite{B+R}). The algebraic quantization method (\cite{bluebook})
proceeds in a different (but more or less equivalent) way: the relations
generate an ideal \(I_R\) in the algebra
\(A\); then, one is to look for the representations of the algebra \(A/I_R\).
We must use a different procedure, because we are looking for a representation
of a group (and the group structure does not contain information about
relations); our proceedure can be quite practical, however:
the relations that lie in the center of the algebra can give the Casimir
operators of the group some definite values. Then, the physical representation
is determined or limited strongly (for examples, see \cite{pohlmayer}).

\section{Time evolution}
\label{sec:evol}
In this section, we will generalize the construction of the time evolution as
described in \cite{timelevels}. The key idea in \cite{timelevels} is to
introduce an auxiliary rest
frame in the phase space and to describe the movement of the system with
respect to this frame. The rest frame is constructed in such a way that the
resulting time evolution reproduces the usual results for parametrized systems
with well-known time evolution.

\subsection{General theory}
\label{sec:general}
Let \(\{\Gamma_t\}\) be a family of
transversal surfaces and \(t\in{\cal T}\), where \({\cal T}\) is an index set
(it can contain just two elements, it can coincide with the real axis, etc.).
There is a symplectic form \(\Omega_t\) associated
with each \(t\) as described in section \ref{sec:singular}. Thus, we have
the symplectic manifolds \((\Gamma_t ,\Omega_t )\), which will be called {\em
time levels}. Let \(\vartheta_{tt'} : \Gamma_t\mapsto\Gamma_{t'}\) be a
symplectic diffeomorphism for each pair \((t,t')\); this maps will be called
{\em time shifts}. Finally, the system \(\{\Gamma_t ,\vartheta_{tt'}\}\) is
called
{\em auxiliary rest frame}. A dynamical trajectory of the system with respect
of the
auxiliary rest frame can be defined as follows. Let \(\gamma\) be a c-orbit (a
maximal classical solution in all possible gauges and foliations). Suppose
that \(\gamma\cap\Gamma_t \neq\emptyset\) for all \(t\in{\cal T}\). Then
\(\gamma\) determines a map \(\eta_{\gamma}:{\cal T}\mapsto\Gamma\) by
\[ \eta_{\gamma}(t)=\gamma\cap\Gamma_t\quad \forall t\in{\cal T}, \]
and this map will be called {\em dynamical trajectory}.

	The time shifts define what might be intuitively
described as ``the same measurements at different times''. Let
\(o_t\) be a perennial whose value is measurable at the time level
\(\Gamma_t\). Thus, \(o\) is associated with a particular measurement at this
time level (an apparatus in a particular position, etc.). We define the same
measurement
at the time level \(\Gamma_t'\) by the perennial \(\theta_{tt'}o\) that is
given by the relation
\[ (\theta_{tt'}o)|_{\Gamma_t'} = o|_{\Gamma_t}\circ\vartheta_{tt'}^{-1}. \]
(The initial datum for \(\theta_{tt'}o\) at \(\Gamma_{t'}\) is obtained by
mapping that of \(o\) at \(\Gamma_t\) by \(\vartheta_{tt'}\) to
\(\Gamma_{t'}\)). The map \(\theta_{tt'} : C^{\infty}(\Gamma_t )\mapsto
C^{\infty}(\Gamma_{t'})\) is a Poisson algebra homomorphism
(\(\vartheta_{tt'}\) is a symplectic diffeomorphism). We will denote the
results of the time shifting described above by \(o_{t'}\).

	Finally, the time evolution of the system is the
\(t\)-dependence of the results of the same measurements made along the
dynamical trajectory of the system. Thus, it is given by the \(t\)-functions
\(o_t (\eta_{\gamma}(t))\). All this is analogous to the ideas in
\cite{timelevels}, but
the time shifts used in \cite{timelevels} were much more special: they were
defined by a one-dimensional symmetry group.

	The above way of defining the
time shifts seems to be the most general one in the following sense. If we
assume that each measurement at a given time level is represented by a
perennial and that two systems of the same measurements at different time
levels are to be represented by perennials with the same Poisson bracket
algebra, then the time shift must be a symplectic diffeomorphism between the
two time levels. This follows from the following proposition:
\begin{prop}
Let \((\Gamma_1,\Omega_1)\) and \((\Gamma_2,\Omega_2)\) be two symplectic
manifolds, \(\vartheta:\Gamma_1\mapsto\Gamma_2\) a diffeomorphism, \(S_1\) a
set of functions which separates points at \(\Gamma_1\), and let \(\vartheta\)
preserve the Poisson brackets,
\[	\{f,g\}_1 = \{f\circ\vartheta^{-1},g\circ\vartheta^{-1}_2\}_2 \]
for any two elements \(f\) and \(g\) of \(S_1\). Then,
\[	\Omega_1 = \vartheta^*\Omega_2.  \]
\end{prop}
{\bf Proof}
If \(S\) is the set of functions that separates points of a manifold
\(\Gamma\), then the differentials of the elements of \(S\) span
\(T_p^*\Gamma\) almost everywhere in \(\Gamma\). Indeed,
suppose that there is an open set \(U\) in
\(\Gamma\) such that the linear span \(D_p\Gamma\) of the differentials of all
elements of \(S\) at each point \(p\in U\) is a proper subspace of
\(T_p^*\Gamma\). Then there is a vector field \(X\) in \(U\) such that
\(\langle X,df\rangle = 0\) for all \(f\in S\) and each \(p\in U\), because
all differentials are smooth forms. As a consequence, \(f\) is constant along
any integral curve of \(X\) in \(U\) for any \(f\in S\). However, then \(S\)
does not separate points of the curve. Next,
let \(\Omega\) and \(\Omega'\) be two symplectic forms on \(\Gamma\); if \(S\)
is a set of functions whose differentials span \(T_p^*\) at \(p\in \Gamma\);
and if \(\{f,g\}_p=\{f,g\}'_p\) for all functions \(f\) and \(g\) from \(S\) at
\(p\), then \(\Omega (p)=\Omega'(p)\). Indeed,
let \(J:T_p^*\Gamma\mapsto T_p\Gamma\) be defined
by \(\langle\xi ,X\rangle = \Omega(X,J(\xi ))\); \(J\) is a linear isomorphism.
Define \(\Omega^{-1}:T_p\Gamma\times T_p\Gamma\mapsto\mbox{\bf R}\) by
\(\Omega^{-1}(X,Y) = \Omega (J^{-1}X,J^{-1}Y)\). \(\Omega^{-1}\) is a
non-degenerated skew-symmetric two-form on \(T_p\Gamma\times T_p\Gamma\),
uniquely determined by \(\Omega\) and satisfying the relation
\(\{f,g\}=\Omega^{-1}(df,dg)\). Similarly, \(\{f,g\}'=\Omega^{\prime
-1}(df,dg)\). It follows that \(\Omega^{-1}=\Omega^{\prime -1}\) and
this implies that \(\Omega = \Omega'\). Finally,
let \(\Omega_2' = \vartheta^{-1*}\Omega_1\);
\(\Omega_2'\) and \(\Omega_2\) are two symplectic forms on \(\Gamma_2\).
Define \(S_2\) by \(S_2 = \{f\in C^{\infty}(\Gamma_2,\mbox{\bf
R})|f=f_1\circ\vartheta^{-1},\ f_1\in S_1\}\). As \(\vartheta\) is a
diffeomorphism, \(S_2\) separates points on \(\Gamma_2\). Hence, \(\Omega_2' =
\Omega_2\) on a dense subset of \(\Gamma_2\). However, \(\Omega_2'\) and
\(\Omega_2\) are smooth. Thus, they are equal everywhere on \(\Gamma_2\), QED.

	The next task is to calculate the numbers \(o_t (\eta_{\gamma}(t))\).
For this purpose,
the information represented by the two \(t\)-functions \(o_t\) and
\(\eta_{\gamma}(t)\) is somewhat superfluous and we are lead to the
Schroedinger and Heisenberg pictures of dynamics (within the classical
theory). In general, the Heisenberg phase space \((\Gamma_H,\Omega_H)\) will
not be the same as the
Schroedinger one \((\Gamma_S,\Omega_S)\).

	To construct \((\Gamma_S,\Omega_S)\), we consider
the set \({\bf \Gamma} := \bigcup_{t \in {\cal T}}\{\Gamma_t\}\) and the
eqivalence relation \(\sim_S\) on \({\bf \Gamma}\) defined as follows: \(p
\sim_S q\) if there is \((t, t') \in {\cal T} \times {\cal T}\) such that \(q
= \vartheta_{tt'}p\). Then, \(\Gamma_S := {\bf \Gamma}/\sim_S\).
As \(\vartheta_{tt'}\) is a
diffeomorphism between \(\Gamma_t\) and \(\Gamma_{t'}\),
\(\Gamma_S\) is diffeomorphic to any of \(\Gamma_t\)'s. As
\(\vartheta_{tt'}\) is symplectic map, \(\Omega_S\) is well-defined on
\(\Gamma_S\). The class \(\{o_t\}:=\{o_t|t\in{\cal T}\}\) of perennials
defines a unique
function on \(\Gamma_S\), as \(o_t\) and \(o_{t'}\) are related by
the pasting \(\vartheta_{tt'}\); let us denote this function by \(o_S\)
and call it {\em Schroedinger observable}. Any dynamical trajectory
\(\eta_{\gamma}\) defines the map \(\eta_{\gamma}^S : {\cal
T}\mapsto\Gamma_S\); this will be called {\em Schroedinger trajectory}
of the system. We obtain easily that
\[ o_t (\eta_{\gamma}(t)) = o_S (\eta_{\gamma}^S (t)). \]

	For the construction of \((\Gamma_H,\Omega_H)\), the procedure is
analogous, but the relation \(\sim_H\) is defined by the maps \(\rho_{tt'} :
\Gamma_t\mapsto\Gamma_{t'}\) where
\(\rho_{tt'} = \pi_{t'}|_{\Gamma_t}\) and \(\pi_{t'}\) is the projector
associated with transversal surface \(\Gamma_{t'}\). The resulting
manifold \(\Gamma_H\) is not necessarily Hausdorff. If the domains of all
\(\Gamma_t\)'s cover \(\Gamma\), then \(\Gamma_H\)
coincides with the quotient space \(\Gamma /\gamma\). Again, there is a
well-defined symplectic form \(\Omega_H\), because the maps
\(\rho_{tt'}\) are symplectic (see  \cite{timelevels}).
A dynamical trajectory \(\eta_{\gamma}\) defines a unique point
\(\eta_{\gamma}^H\) on \(\Gamma_H\) as \(\rho_{tt'}\) acts along
the c-orbits. Any perennial \(o'\)
defines a function, \(o_H\), on \(\Gamma_H\), as it is
invariant with respect to \(\rho_{tt'}\). Thus, the perennials \(o_t\) define
the set of functions \(o_t^H :=o_{tH}\), which is called {\em Heisenberg
observable}. We easily find that
\[ o_t (\eta_{\gamma}(t)) = o_t^H (\eta_{\gamma}^H ). \]

	We observe that the whole class \(\{o_t\}\) of perennials collapses
into one observable of Schroedinger or Heisenberg type. This gives us the
motivation to call such classes {\em observables}. In fact, a proposal to
distinguish between observables and perennials is not new. It has been made
by Kucha\v{r} \cite{cordoba}. His proposal is, however, not equivalent to
ours, because Kucha\v{r} defines observables in a different way. Some
discussion of these and related questions is contained in the section
\ref{sec:peren}.

	The construction of quantum evolution follows closely the classical
one. Let us assume in this subsection that all transversal surfaces
are global; for the modifications due to non-global surfaces, see
subsection \ref{sec:nonglob}. Then, we obtain that
\((\Gamma_S, \Omega_S) \cong (\Gamma_H, \Omega_H) \cong (\Gamma_0,
\Omega_0)\), where
\(0\) symbolizes a fixed element of \({\cal T}\) and \(\cong\) is the
isomorphism of symplectic manifolds. From the definition of \(\theta_{ts}\) it
follows immediately that \(o^H_t=\theta_{0t}o^S=\theta_{st}o^H_s\). The maps
\(\theta_{st}\) have the physical meaning of time evolution maps for the
classical Heisenberg picture; they have to be taken over into
the quantum theory. Recall that we have the representation \(R:S\mapsto
L(K)\) already at our disposal. If \(\theta_{st}S\subset S\) (this happens in
linear theories, like
quantum field theory on curved background, cf. \cite{H+I}), then it is
straightforward to define \(\hat{\theta}_{st}\) by the commuting diagram:
\[
\begin{array}{ccc}
	S & \stackrel{\theta_{st}}{\longrightarrow} & S \\
	\downarrow R & & \downarrow R \\
	L(K) & \stackrel{\hat{\theta}_{st}}{\longrightarrow} & L(K)
\end{array}.
\]
In the oposite case, one has to choose one element of the algebra \(\bar{A}\)
for each \(o\in S\), \(s\in {\cal T}\) and \(t\in{\cal T}\)
to play the role of \(\theta_{st}(o)\) (which is an element of \(\cal P\);
this is another factor ordering problem). The result would be a map
\(\theta^a_{st}:S\mapsto \bar{A}\). There are
some reasonable restriction on this map \(\theta^a_{st}\) (or else the choice
is practically unlimited!): we require the following two conditions:
\begin{enumerate}
\item \(\theta^a_{st}\) is a Lie-algebra isomorphism of \(S\) and
\(\theta_{st}^a(S)\),
\item \(\theta^a_{st} = \theta^a_{ut}\circ\theta^a_{su}\) for all \(s,\ t\) und
\(u\) for which the equation \(\theta_{st} = \theta_{ut}\circ\theta_{su}\) is
satisfied.
\end{enumerate}
Then the corresponding quantum map \(\hat{\theta}_{st}\) is defined by the
following diagram
\[
\begin{array}{ccc}
	S & \stackrel{\theta^a_{st}}{\longrightarrow} & \bar{A} \\
	\downarrow R & & \downarrow R \\
	L(K) & \stackrel{\hat{\theta}_{st}}{\longrightarrow} & L(K)
\end{array},
\]
because the representation \(R\) can be extended to the algebra \(\bar{A}\).

	Having the quantum map \(\hat{\theta}_{st}\), we can attempt to
implement it by a unitary map \(U_{st}:K\mapsto K\) so that
\(\hat{\theta}_{st}(\hat{o}) = U_{st}\hat{o}U_{st}^{-1}\) for \(\hat{o}\in
L(K)\). \(\{U_{st}\}\) is the system of unitary evolution operators for the
system, and the construction of the (quantum) Schroedinger and Heisenberg
picture can be completed in a straightforward way.

	We will clarify and develop the general concepts as introduced in this
section by studying some particular cases.

\subsection{Continuous, one dimensional case}
\label{sec:cont}
For the sake of simplicity, we will assume in
this subsection that all transversal surfaces are global. This assumption can
easily be removed by working within the domain of a non-global surface.

	Let \(\{\Gamma_t\}\) be a one-dimensional differentiable family of
global transversal surfaces; then \({\cal T}={\bf\mbox{R}}\). Let
\(\vartheta_t : \Gamma_0\mapsto\Gamma_t\) be a symplectic diffeomorphism for
each \(t\) such that \(\vartheta_t (p)\) is a smooth curve for each
\(p\in\Gamma_0\); these curves define the ``rest trajectories''.
Each ordered pair of time levels defines the time shift by
\(\vartheta_{tt'} = \vartheta_{t'}\circ\vartheta_t^{-1}\). Any dynamical
trajectory is a curve \(\eta_{\gamma}:\mbox{\bf R}\mapsto\Gamma\).
This curve is a classical solution in a particular gauge and foliation; this
is why it is a one-dimensional object. A perennial \(o\) measurable at
\(\Gamma_0\) defines an observable \(\{o_t\}\) as described in section
\ref{sec:general}.

	The Schroedinger phase space is isomomorphic to \((\Gamma_0 ,\Omega_0
)\), the Schroedinger trajectory \(\eta_{\gamma}^S(t)\) is obtained by
\(\eta_{\gamma}^S(t)=\theta_t^{-1}\eta_{\gamma}(t)\) and the Schroedinger
observable is given by \(o_S=\theta_t^{-1}(o_t )=o_0\). The Heisenberg
phase space is also isomorph to \((\Gamma_0 ,\Omega_0)\) as the maps
\(\rho_{tt'}\) are symplectic diffeomorphisms. Each c-orbit \(\gamma\) defines
a Heisenberg trajectory, the point \(\eta_{\gamma}^H=\gamma\cap\Gamma_0\). Each
observable \(\{o_t\}\) defines the Heisenberg observable \(o_t^H\) on
\(\Gamma_0\) by projecting each perennial \(o_t\) to \(\Gamma_0\):
\(o_t^H = o_t |_{\Gamma_0}\). The \(t\)-functions \(\eta_{\gamma}^S(t)\) and
\(o_t^H\) satisfy ordinary differential equations, which we are going to
derive.

	The Schroedinger trajectories define a set of maps
\(\chi_{tt'}:\Gamma_0\mapsto\Gamma_0\) on the Schroe\-din\-ger phase space as
follows. Let \(p\in\Gamma_0\) and \((t,t')\in\mbox{\bf R}^2\);
then
\be
\chi_{tt'}:=\theta_{t'}^{-1}\circ\rho_{0t'}\circ\rho_{0t}^{-1}\circ\theta_t .
\label{composchi}
\ee
{}From this definition, it follows directly:
\begin{enumerate}
\item the relation
\[
 	\eta_{\gamma}^S(t') = \chi_{tt'}(\eta_{\gamma}^S(t)),
\]
\item that \(\chi_{tt'}\) is a symplectic diffeomorphism for each
	\((t,t')\in\mbox{\bf R}^2\),
\item the composition law \(\chi_{tt'} = \chi_{st'}\circ\chi_{ts}\) for all
	\((t,t',s)\in\mbox{\bf R}^3\).
\end{enumerate}
In particular, \(\chi_{ts}^{-1}=\chi_{st}\). However,
\(\chi_{st}\neq\chi_{0(t-s)}\), in general, i.\ e., the set of maps
\(\chi_{0t}\) does not form a group (it is no flow!).

	It is easy to prove that \(\dot{\eta}_{S\gamma}(t)=X_t^S
(\eta_{\gamma}^S(t))\), where \(X_t^S\) is a locally Hamiltonian vector field,
as \(\chi_{st}\) is symplectic and because of Eq.\ (\ref{composchi}). Suppose
that \(X_t^S\) is globally Hamiltonian; let us call the auxiliary rest frame
{\em Hamiltonian} in this case. Then there is a function \(H_t^S
:\Gamma_0\mapsto\mbox{\bf R}\) such that \(X_t^S\) is its Hamiltonian vector
field, and we have
\be
	\dot{\eta}^S_{\gamma}(t) = X_t^S .
\label{Sdyneq}
\ee
The set of functions \(\{H_t^S |t\in\mbox{\bf R}\}\) is called {\em
Schroedinger Hamiltonian} and Eq. (\ref{Sdyneq}) is the {\em Schroedinger
equation of motion}.

	For the Heisenberg observable \(o_t^H\), we have that \(o_t^H(p)=o_t
(\rho_{0t}(p))\) and \(o_t (\rho_{0t}(p)) = o_0
((\theta_t^{-1}\circ\rho_{0t}(p))\). Thus,
\be
	o_t^H = o_S\circ\chi_{0t},
\label{SH}
\ee
because \(o_0 = o_S\). Let us calculate the derivative of \(o_t^H\) with
respect to \(t\); using Eq. (\ref{SH}), we obtain:
\be \dot{o}_t^H(p) = (X_t^S\cdot o_S )(\chi_{0t}(p))=(\chi_{0t\,*}^{-1}
X_t^S )(o_s\circ\chi_{0t})|_p = X_t^H\cdot o_t^H,
\label{odot}
\ee
where \(X\cdot f\) is the action of the vector field \(X\) (as a
differential operator) on the function \(f\) and \(X_t^H =
\chi_{0t\,*}^{-1}X_t^S\);
as \(X_t^S\) is a Hamiltonian vector field of \(H_t^S\) and \(\chi_{0t}\) is a
symplectic diffeomorphism, \(X_t^H\) is a Hamiltonian vector field of the
function
\be
	H_t^H = H_t^S\circ\chi_{0t}.
\label{SHhamilt}
\ee
The set of functions \(\{H_t^H |t\in\mbox{\bf R}\}\) is called {\em Heisenberg
Hamiltonian} and Eq. (\ref{odot}) implies the {\em Heisenberg equation of
motion}
\be
    \dot{o}^H_t = \{o^H_t,H^H_t\}.
\label{Heq}
\ee

	Now, we can return to the discussion of the relation between integrals
of motion and perennials. Clearly, each perennial \(o\) defines a function on
the reduced phase space \(\Gamma_0\) by \(i_0^*o\). Eq. (\ref{Heq}) shows that
\(i_0^*o\) is an integral of motion, if the time shifts are chosen so as to
preserve \(o\). Thus, any given perennial can become an integral for some time
evolution.

\subsubsection{An example}
\label{sec:example}
Let \(\tilde{\Gamma}\) be \(\mbox{\bf R}^{2n+2}\) with canonical coordinates
\(T,P,q^1 ,\ldots ,q^n ,p_1 ,\ldots ,p_n\) and let \(\tilde{\Omega}=dP\wedge
dT+ dp_k\wedge dq^k\). The constraint surface is
given by the equation \(C=0\), where \(C\) is a differentiable function on
\(\tilde{\Gamma}\); let the equation \(C=0\) be equivalent to
\be
	P=-{\cal H}(T,q^1 ,\ldots ,q^n ,p_1 ,\ldots ,p_n ),
\label{constraint}
\ee
where \({\cal H}\) is a smooth function on \(\mbox{\bf R}^{2n+1}\). This
defines our first-class parametrized system.

	We choose an auxiliary rest frame as follows. Let \(\Gamma_t\) be
the image of embedding \(\mbox{\bf R}^{2n}\) with canonical coordinates \(x^1
,\ldots ,x^n ,y_1 ,\ldots ,y_n \) by the embedding maps \(i_t\) into
\(\tilde{\Gamma}\) that are given by
\[ i_t (x^1 ,\ldots ,x^n ,y_1 ,\ldots ,y_n)=(t,-{\cal H}(t,x^1 ,\ldots ,x^n
,y_1 ,\ldots ,y_n ),x^1 ,\ldots ,x^n ,y_1 ,\ldots ,y_n ). \]
Observe that \({\cal H}(x^1 ,\ldots ,x^n ,y_1 ,\ldots ,y_n )=-P|_{\Gamma_t}\).
Clearly, \(\Omega_t = dy_k\wedge dx^k\).
Let the maps \(\vartheta_t\) be given by \(\vartheta_t=i_t\circ i_0^{-1}\).

	A tangent vector field \(L\) to the c-orbits is easily calculated from
the constraint in the form (\ref{constraint}):
\[
	L=\frac{\partial}{\partial T} - \frac{\partial{\cal H}}{\partial T}
	\frac{\partial}{\partial P} + \frac{\partial{\cal H}}{\partial p_k}
	\frac{\partial}{\partial q^k} -\frac{\partial{\cal H}}{\partial q^k}
	\frac{\partial}{\partial p_k},
\]
For the Schroedinger dynamical trajectory, we obtain simply that
\(\eta_{\gamma}^S(t)=\vartheta_t^{-1}(\eta_{\gamma}(t))\); thus,
the tangent vector \(X_t^S\) to this trajectory is given by
\(X_t^S (p)=\vartheta_{t*}^{-1}L(\vartheta_t(p))\), which results in
\[
	X_t^S = \left(\frac{\partial{\cal H}}{\partial y_k}
	\frac{\partial}{\partial x^k} -\frac{\partial{\cal H}}{\partial x^k}
	\frac{\partial}{\partial y_k}\right)_{T=t}.
\]
It follows that the Schroedinger Hamiltonian is
\[
	H^S (t,x^1 ,\ldots ,x^n ,y_1 ,\ldots ,y_n ) = \vartheta_t^*{\cal
	H}|_{T=t}.
\]
Observe that the family of rest trajectories is not generated by \(P\) in
general.

	Next, we define the perennials \(Q^k_t\) and \(P_{tk}\) by
\[  Q^k_t|_{\Gamma_t} = x^k,\quad P_{tk}|_{\Gamma_t} = y_k. \]
Clearly, \(Q_s^k = \theta_{ts}Q_t^k\) and \(P_{sk} = \theta_{ts}P_{tk}\) for
any pair \((t,s)\) of real numbers. For each value of \(t\), we obtain a
complete system of perennials with a well-known algebra.

	The Heisenberg Hamiltonian \(H^H (p)=H^S (\chi_{0t}(p))\) is not
available in explicit form, as \(\chi_{0t}\) can only be obtained by
integrating the differential equation \(d\chi_{0t}/dt=X_t^S\).

	The procedure desribed in this subsection is, on one
hand, equivalent to that in \cite{timelevels}, if the time shifts
\(\vartheta_t\) are generated by a perennial \(h\); then
\(H^S=H^H=-h|_{\Gamma_0}\). On the
other hand, the example shows that it is related to the so-called reduction
procedure, which is the reversal of the parametrization procedure that was
described in section \ref{sec:singular}, see also \cite{I+B} or
\cite{moncrief}.

\subsection{Non-global transversal surfaces}
\label{sec:nonglob}
Consider the following situation. Let \(\Gamma_1\) and \(\Gamma_2\) be two
transversal surfaces; let \( \Gamma''_1 :=\Gamma_1\cap{\cal D}(\Gamma_2 )\) and
\( \Gamma'_2 :=\Gamma_2\cap{\cal D}(\Gamma_1 )\); let \(\rho
:\Gamma''_1\mapsto\Gamma'_2\) be given by \(\rho
:=\pi_2'|_{\Gamma''_1}\), where \(\pi_2'\) is the projector associated with
\(\Gamma_2'\); and finally let
\(\vartheta :
\Gamma_1\mapsto\Gamma_2\) be a time shift (\(\rho\) and \(\vartheta\) are
symplectic diffemorphisms). Our aim is to construct a quantum time evolution
from the time level \(\Gamma_1\) to \(\Gamma_2\) in the general case when
\(\Gamma''_1\) and \(\Gamma'_2\)
are proper submanifolds of \(\Gamma_1\) and \(\Gamma_2\).

	In order to get some hint of how one can proceed and how the problem
is to be posed, let us stay within the classical theory and consider an
evolution of an ensemble of classical systems on \(\Gamma\); let this
ensemble be described by  a measure \(\mu\) on \(\Gamma/\gamma\); let
\(\mu_1\) and \(\mu_2\) be the measures induced by \(\mu\) on \(\Gamma_1\) and
\(\Gamma_2\), respectively.

	The problem can now be posed as follows. Suppose that we can control
the input only at \(\Gamma_1\), and that we can measure the output only at
\(\Gamma_2\). In particular, we can prepare the \(\Gamma_1\)-part of the
ensemble arbitrarily so that \(\mu_1\) can be normalized,
\(\int_{\Gamma_1}d\mu_1 = 1\). Which perennials have then a mean value at
\(\Gamma_2\)  that is calculable from the knowledge of \(\mu_1\)? Let us first
study a simpler question: Suppose that a transversal surface \(\Gamma_0\),
not necessarily global, and the measure \(\mu\) are given. Which perennial has
a mean value calculable from what is known at \(\Gamma_0\)?
The problem is that the data at \(\Gamma_0\) do not determine the measure
outside of \({\cal D}(\Gamma_0 )/\gamma\) so that the mean value of a
perennial that does not vanish there is not determined.
This leads to the following definition.
\begin{df}
	The perennial \(o\) is called {\em pertinent} to the transversal
surface \(\Gamma_0\), if
\begin{enumerate}
\item \(o(\Gamma\setminus{\cal D}(\Gamma_0 ))=0\),
\item the Hamiltonian vector field \(\xi_o\) of \(o\) is complete on \({\cal
	D}(\Gamma_0 )\).
\end{enumerate}
\end{df}
Then clearly,
\be
	\mbox{mean}(o)=\int_{\Gamma_0}d\mu_1\,o.
\label{mean}
\ee
For example, if the perennial
\(o\) generates a symmetry group which leaves \({\cal D}(\Gamma_0 )\)
invariant, then \(\Pi_{\Gamma_0}o\) is pertinent to \(\Gamma_0\). The condition
1 is sufficient for Eq. (\ref{mean}) to hold, but the condition 2
will turn out to be vital for the quantum theory.

	If \(o_1\) is pertinent to \(\Gamma_1\) then \(o_2 :=\theta o_1\) is
pertinent to \(\Gamma_2\). The pair \((o_1 ,o_2 )\) is an observable
associated with the auxiliary rest frame \((\Gamma_1 ,\Gamma_2 ,\vartheta)\).

	Let us study the evolution of the mean values of observables. Let
\((o_1 ,o_2 )\) be an observable; we want to calculate \(\mbox{mean}(o_2 )\)
using only \(\mu_1\), \(o_1\), \(\rho\) and \(\vartheta\). \(o_2\) is
determined everywhere at \(\Gamma_2\) by these data, but \(\mu_2\) is
determined only at \(\Gamma'_2\):
\[ \mu_2|_{\Gamma'_2}=\rho^{-1*} (\mu_1 ); \]
the rest \(\mu_2|_{\Gamma_2\setminus\Gamma'_2}\) of \(\mu_2\), which is not
controlled from \(\Gamma_1\), can be considered as {\em noise}; we assume that
it is completely independent of \(\mu_1\). Thus, \(o_2\) must be pertinent to
\(\Gamma'_2\) and \(o_1\) to \(\Gamma'_1:=\vartheta^{-1}(\Gamma'_2
)\).

	Let \({\cal P}_{\Gamma'_2}\) and \({\cal
P}_{\Gamma'_1}\) be the algebras of all perennials that are
pertinent to \(\Gamma'_2\) and \(\Gamma'_1\), respectively.
We have:
\[ \theta{\cal P}_{\Gamma'_1 }= {\cal P}_{\Gamma'_2}. \]
Moreover,
\[ {\cal P}_{\Gamma'_2} = {\cal P}_{\Gamma''_1}\subset{\cal P}_{\Gamma_1}. \]
Thus, \({\cal P}_{\Gamma'_2}\) is determined by generators of
\({\cal P}_{\Gamma_1}\) (even by those of the subalgebra \({\cal
P}_{\Gamma''_1}\) of \({\cal P}_{\Gamma_1}\).

	In the (classical) Schroedinger picture, the two transversal surfaces
\(\Gamma_1\) and \(\Gamma_2\) are identified by \(\vartheta\) to, say,
\(\Gamma_1\). \(\Gamma''_1\) remains as it is and \(\Gamma'_2\) becomes to
\(\Gamma'_1\). The map \(\chi : \Gamma''_1\mapsto\Gamma'_1\) was defined
in the subsection \ref{sec:general} by \(\chi = \vartheta^{-1}\circ\rho\).
Let \((o_1 ,o_2 )\) be an observable such that \(o_2\) pertains to
\(\Gamma'_2\). Then it holds that
\[ \int_{\Gamma_2}d\mu_2\,o_2 = \int_{\Gamma_1}d(\chi^{-1*}\mu_1 )o_1 . \]
Thus, the evolution is given by the map \(\chi^{-1*}\) of \(\mu_1\).

	In the (classical) Heisenberg picture, we identify \(\Gamma_1\) and
\(\Gamma_2\) by \(\rho\) along \(\Gamma''_1\) and \(\Gamma'_2\). There is only
one measure, \(\mu\). Only the time shifts of the observables from \({\cal
P}_{\Gamma'_1}\) screen the noise authomatically; their images by the time
shift \(\theta\) lie in \({\cal
P}_{\Gamma'_2}\). The mean values calculated in the Schroedinger picture
coincide with the corresponding Heisenberg picture ones.

	The analysis above suggests that the following groups and algebras
will play an important role in the  quantization. Let \(G_1\) and \(G_2\) be
two groups of symplectic diffeomorphisms acting on \(\Gamma_1\) and
\(\Gamma_2\), respectively, and let \(S_1\) and \(S_2\) be the Lie algebras of
functions on \(\Gamma_1\) and \(\Gamma_2\) that generate these groups via
Poisson brackets. The groups \(G_1\) and \(G_2\) may result as projections to
\(\Gamma_1\) and \(\Gamma_2\) of some groups of symmetries in
\(\tilde{\Gamma}\). The functions from \(S_1\) and \(S_2\) define perennials
with the same algebras and
we will denote these algebras of perennials by the same symbols. Let the groups
\(G_1\) and \(G_2\) satisfy the following requirements:
\begin{enumerate}
\item \(G_2=\{\vartheta\circ g\circ\vartheta^{-1}|g\in G_1\}\),
\item \(S_2=\{o\circ\vartheta^{-1}|o\in S_1\}\).
\end{enumerate}
Thus, the groups \(G_1\) and \(G_2\) are isomorphic, and their actions are
related by \(\vartheta\). Let \(G'_2\in G_2\) be the subgroup which preserves
\(\Gamma'_2\). \(G_2'\) acts on \(\Gamma'_2\in\Gamma_2\), but it has
also an action \(a_1\) on \(\Gamma''_1\), because it
preserves the common domain of \(\Gamma'_2\) and \(\Gamma''_1\); \(a_1\) is
defined by \(a_1(g):=\rho^{-1}\circ g\circ\rho\) for all \(g\in G'_2\). Thus,
\(a_1(G'_2)\) acts on \(\Gamma_1\), but it is no subgroup of \(G_1\), in
general. The algebra of perennials that generate \(G'_2\) will be denoted by
\(S'_2\). The projections of the perennials from \(S'_2\) generate the
action of \(G'_2\) on both \(\Gamma'_2\) and \(\Gamma''_1\). Finally,
\(G'_1\) is the subgroup of \(G_1\) which is related by \(\vartheta\) to
\(G'_2\), that is \(G'_1:=\{\vartheta^{-1}\circ g\circ\vartheta|g\in G'_2\}\).
Then \(G'_1\) preserves \(\Gamma'_1\). It follows that each element \(g_2\in
G'_2\) defines an element \(g_1\in G'_1\) such that
\(a_1(g_2) = \rho^{-1}\circ\vartheta\circ g_1\circ\vartheta^{-1}\circ\rho =
\chi^{-1}\circ g_1\circ\chi\). Let \(S'_1\) be the algebra of perennials that
generates \(G'_1\); then \(S'_2 = \theta S'_1\). For the projections of the
algebras, we obtain easily \(S'_2|_{\Gamma_1} = \{o_1\circ\chi^{-1}|o_1\in
S'_1|_{\Gamma_1}\}\).

	As the elements of \(\theta S'_1\) do not lie in \(S_1''\) in general,
we have to look for them in the universal enveloping algebra \(\bar{A}_1\) of
\(S_1\). Suppose that we have solved this ``factor-ordering problem''. Let
us denote by \(\theta_a(o)\) the element of \(\bar{A}_1\) which is associated
in this way with \(o\in S'_1\).

	The construction of the corresponding quantum mechanical evolution
is based on an analogous problem setting: if we can prepare a state at the
time level \(\Gamma_1\), what can be said about measurements at the time level
\(\Gamma_2\)? The answer can be worked out with the tools of the ordinary
quantum mechanics and it consists of the following steps.
\begin{enumerate}
\item  	With the two phase spaces \(\Gamma_1\) and \(\Gamma_2\), we associate
	the Hilbert spaces \(K_1\) and \(K_2\) and the representations \(R_1\)
	and \(R_2\) of the groups and algebras, \(R_1:G_1\mapsto L(K_1)\) and
	\(R_2:G_2\mapsto L(K_2)\). These are unitarily equivalent, irreducible
	unitary representations, and let the unitary equivalence be realized
	by the map \(U(\vartheta):K_1\mapsto K_2\). In most cases, one just
	takes two copies of the same representation, so the search for
	\(U(\vartheta)\) is trivial.
\item 	We try to find the Hilbert subspaces that correspond to the symplectic
	ma\-ni\-folds \(\Gamma'_1\), \(\Gamma''_1\) and \(\Gamma'_2\) using
	the method described in \cite{patchI}. Consider \(\Gamma'_1\).
	The representation \(R_1:G'_1\mapsto L(K_1)\) is not (in general)
	irreducible. Thus, \(K_1\) decomposes into irreducible
	representations subspaces. Each such subspace is usually characterized
	by values of invariants (in particular, the Casimirs elements) of
	\(G'_1\). A comparison with the classical values of these invariants
	on \(\Gamma'_1\) helps to identify the subspace \(K'_1\) that
	corresponds to the classical submanifold \(\Gamma'_1\) in the quantum
	mechanics. Similarly for \(\Gamma''_1\) and \(\Gamma'_2\) we find the
	subspaces \(K''_1\subset K_1\) and \(K'_2\subset K_2\). Clearly,
	\(K'_2=U(\vartheta)K'_1\).
\item	The construction of the Schroedinger picture proceeds by identifying
	the Hil\-bert spaces \(K_1\) and \(K_2\)
	using the map \(U(\vartheta )\). Then the map \(\chi :
	\Gamma''_1\mapsto\Gamma'_1\) is to be implemented by a unitary map
	\(U(\chi ):K''_1\mapsto K'_1\). \(\chi\) is a
	symplectic diffeomorphism with domain \(\Gamma''_1\) that may be
	singular at the boundary \(\partial\Gamma''_1\). One possible method
	is to look for a function \(h\) on \(\Gamma_1\) (it may be singular
	at the boundary) that generates a flow such that the map \(\chi\) is
	the element of
	the flow at the value 1 of the flow parameter. Then the factor order
	problem has to be solved: the function \(h\) is to be identified with
	an element \(h_a\) of \(\bar{A}_1\). Finally, we set \(U(\chi ) = \exp
	(R_1 h_a)|_{K''_1}\). An example in which this method works is given in
	section \ref{sec:patch}. The dynamics in the Schroedinger picture is
	given by \(U(\chi )\) in the Schroedinger Hilbert space
	\(K_1\). As such, it is not a unitary map in general: formally,
	neither its domain nor its range coincide with \(K_1\); less formally,
	the evolution of the state \(\psi\in K_1\) is given by \(U(\chi
	)P'_1\psi\), where \(P'_1\) is the projection operator on the subspace
	\(K'_1\) of \(K_1\). Thus, if \(\psi\) has norm \(1\), its time
	evolution will have norm \(\leq 1\).
\item	To construct the Heisenberg picture, we have to identify the Hilbert
	spaces \(K_1\) and \(K_2\) along the subspaces \(K''_1\) and \(K'_2\)
	using an implementation of the symplectic diffeomorphism
	\(\rho:\Gamma''_1\mapsto\Gamma'_2\). As \(\rho = \vartheta\circ\chi\),
	we can set \(U(\rho )=U(\vartheta )\circ U(\chi )\) utilizing our
	knowledge of the map \(U(\chi )\). An alternative way is to define the
	map \(\hat{\theta}\) by the commuting diagram
	\[
	\begin{array}{ccc}
	  S'_1 & \stackrel{\theta_a}{\longrightarrow} & \bar{A}_1 \\
	  \downarrow R_1 &          & \downarrow R_1 \\
	  L(K'_1) & \stackrel{\hat{\theta}}{\longrightarrow} & L(K_1)
	\end{array}. \]
	Then we attempt to implement \(\hat{\theta}\) by a unitary
	map \(U(\chi )\) so that \(\hat{\theta}(\hat{o}) = U(\chi
	)\circ\hat{o}\circ U^{-1}(\chi )\) for all \(\hat{o}\in R_1(S'_1)\)
	such that \(\mbox{Dom}(U(\chi )) = K''_1\) and \(\mbox{Ran}(U(\chi ))
	= K'_1\). This may be a problem, because \(\theta_a\) is defined on a
	proper subalgebra of \(S_1\) only. An example in which it works is
	described in section \ref{sec:patch}. Using \(U(\chi )\), one can
	paste the Hilbert spaces as above finishing the construction. In the
	Heisenberg picture, the measurement of the observables from
	\(\hat{\theta}(R_1S'_1)\) is
	predictable, because they leave the subspace \(K'_2=K''_1\) invariant.
	The expansion of any state \(\psi\) of \(K_2\) into the eigenvectors
	of these observables is well-defined even if we know only the
	\(K'_2\)-projection of \(\psi\). This is the main reason behind the
	point 2 of the definition 1. Quantum mechanically, one can equivalently
	require that the elements of \(S'_1\) or \(S''_1\)
	commute with the projectors \(P'_1\) or \(P''_1\); then, one can
	always multiply the elements of \(S'_1\) or \(S''_1\) by the projector
	\(P_1\) or \(P''_1\) obtaining again self-adjoint operators. Thus,
	one of the main features of the  Heisenberg picture---the
	time-independence of the states---can be preserved if we limit
	ourselves to the measurement of just the observables that are
	pertinent to \(K'_2\). If \(G_1'\) or \(G''_1\) do not act
	transitively on \(\Gamma_1'\) or \(\Gamma''_1\) or if \(S_1'\) or
	\(S''_1\) do not separate points in \(\Gamma_1'\) or \(
	\Gamma''_1\), then the system of measurements defined by the
	observables from \(S_1'\) or \(S''_1\) is not complete in \(K_1'\) or
	\(K''_1\) and the genuine Heisenberg picture of a complete quantum
	evolution cannot be constructed. However, one
	can pass to a kind of a mixed picture instead. One can obtain a
	complete information by performing measurements corresponding
	to the elements of the algebra \(S_2\) that is pertinent to the whole
	space \(K_2\), if one can screen away the noise from the states
	by the projection operator \(P_2'\) before these measurements are done
	(for an example of such a case, see section \ref{sec:patch}). Thus,
	the time evolution of the states is given by the projection and that
	of the observables by the map \(U(\vartheta )\) (which coincides with
	\(U(\chi )\) on \(K_1''\), because \(U(\rho )\) is an identity).
\end{enumerate}

	One may be able to find pathological classical systems for which
this construction cannot be performed, but we
hope that it will work in physically interesting cases.

\section{Meaning of perennials}
\label{sec:peren}
The perennial formalism is based on two ideas:
\begin{itemize}
\item	Study the systems whose time evolution is well-understood like
	Newtonian systems \cite{timelevels}, the
	massive particle in Minkowski spacetime \cite{honeff} or the
	scalar field in curved spacetime \cite{H+I}. These systems all posses
	a background spacetime and some structure of this spacetime plays a
	crucial role in the construction of quantum evolution.
\item	Replace this spacetime structure by or transform it into some phase
	space structure so that the quantum time evolution of the systems can
	be reconstructed {\em solely from some phase space objects}. Try to
	use similar phase space objects to construct a quantum time
	evolution for systems without any background spacetime.
\end{itemize}
The approach seems a little formal in comparison with attempts to reconstruct
time by using some physical system playing the role of a clock \cite{clock}
or in which time is to emerge in the semiclassical approximation
\cite{semiclass}. The hope is that we can reconcile our approach with these
attempts (this is a project for future research).

	A key mathematical notion that keeps everything together and allows
elegant proofs and formulations is that of a perennial. The perennial
formalism is a kind of language that is adequate to
describe the relevant structure of parametrized systems. However, there has
been some discussion in the literature about perennials (or about equivalent
notions), cf.\ \cite{unruh}, \cite{cordoba}, \cite{jim}, \cite{MRFrames}, or
\cite{rov+haj}. What is the relation of our perennial formalism to the ideas
that come out of this discussion?

	Two problems were already discussed: that of existence of perennials
(section \ref{sec:singular}), and that of having explicit expressions for
perennials \cite{timelevels}.

	A very important point is the relation between perennials
and observables. A thorough discussion of this relation is given in
\cite{cordoba}. The conclusion was that ``One can observe dynamical variables
which are not perennial, and...Perennials are often difficult to observe.''
The results of the present paper support Kucha\v{r}' opinion in that the
perennials and observables turn out to be two different notions in general.
More precisely, if we are looking for the classical counterparts of {\em
quantum mechanical observables}---which possess the Schroedinger and
Heisenberg forms---then these are definitely not perennials, because some
``time information'' is contained in them (it is an
interesting question to be studied whether or not there are observable
quantities of different kind). We have identified
such observables with classes of perennials, each two elements of which are
time shifted with respect to each other. It seems to follow that perennials
are in principle measurable, but only in relation to a particular instant of
time (in general, to a transversal surface; for systems equipped with a unique
time, to a particular instant of that time): the value of a perennial at a
given time coincides with the value of an observable that contains the
perennial at the time as an element of the corresponding class. This seems to
be a natural consequence of our approach. However, this
touches another controversy. For those who would consider perennials as
exactly analogous to ``gauge-invariant quantites'' of gauge theories, the way
of
their measurement must also be ``gauge-invariant''; that is, it is either not
associated with any time instant (which is, in fact, a particular location at
a ``gauge orbit'') at all, or it can be performed at any time instant
with the same result. This would also apply to the quantum version of the
theory, and for this version, a very interesting counterexample has been
constructed by Kucha\v{r} \cite{counterex}. Suppose that perennials turn to be
observables in quantum theory that are measurable at any time instant
and that the results of such measurements of one and the same perennial
at different time instants are time independent. Consider a set of
non-commuting perennials. Let us perform two measurements of all perennials in
the set, each in a different time order. From the assumptions, it
follows that the two measurements must give the same result. This, however,
contradicts the basic postulates of the quantum theory of measurement.
The counterexample seems to speak in favour of the distinction between
observables and perennials as it results from our theory. The next remark
concerns the nature of observables. The form that the observables obtain in
this paper (namely, classes of perennials) is not the only form possible. They
may be equivalently described in a way that
makes no reference to perennials. An example is provided by the system
studied in the section \ref{sec:example}. There, e.\ g.\ the classes
\(\{Q^k_t\}\) of the perennials \(Q^k_t\) are observables; each such class is
determined by the coordinate function \(x^k\) (assuming the
time foliation as known); the coordinate \(x^k\) would provide such an
equivalent (but non-geometrical) description of the observable. How may such
an object be measurable at all being no ``gauge invariant?'' The old discussion
of this problem is nicely summarized in \cite{MRFrames}. Briefly, a quantity
\(x\) that is not gauge invariant within a given model A can be associated
with another quantity \(y\) of a model B such that \(y\) is gauge invariant
within B and acquires the same (or approximately the same) values as \(x\) in
the same physical situation. The system B contains the system A as a subsystem
together with some auxiliary matter system (``material reference frames'').
For the measurement of \(x\), the coupling of A to the auxiliary matter system
is in any case necessary. Thus, what is measurable in a given model A is
determined by all possible couplings to other models, not just by A itself.

	Finally, there has been some discussion about perennials of a
particular form, namely ``evolving constants of motion'': roughly, such a
perennial is the value of a quantity taken at the hypersurface in the phase
space that  is defined as a level of some other quantity (reference quantity),
see \cite{rov+haj}. One problem
with these perennials is that they are too complicated functions (for general
reference quantities) to be easily representable by quantum operators; they
will be (continuous) functions with diverging derivatives; their Hamiltonian
vector fields will practically never be complete, etc. (It seems also that the
perennial defined e.\ g.\ as the coordinate the system had at 5 o'clock is
measurable only at five o'clock, cf.\ previous paragraph.) A deeper critics of
such quantities is contained in \cite{jim}: a general reference quantity will
often lead to a perennial that describes a dynamically very involved
information so that its time ordering is not well defined. One has to restrict
the reference quantities to the so-called ``good time functions'', etc.
We have to deal with these objections, because the perennial formalism
also uses quantities analogous to the evolving constants---in fact, the
``observables'' are a kind of such evolving constants, and the perennials that
are defined by their ``initial values'' at some transversal surface are
similar to them, too. However, the reference quantity in all these cases is
chosen such that its levels are transversal surfaces. It seems then that it
must be ``a good time function'', but this is still to be studied in more
detail.

\section{A system without global transversal surfaces}
\label{sec:patch}

\subsection{The model}
\label{sec:model}
An example of a system that did not admit global transversal surfaces was
studied in \cite{patchI}. This system possessed, however, connected
transversal surfaces
that were almost global: their domains were dense in the constraint surface.
The quantum theory of this system did not exhibit, however, much consequence
of the complicated topology of the classical model; this could be shown in
\cite{patchI}. In the present paper, we will give a more interesting example:
there will be inextensible connected transversal surfaces whose domains will
be ``small'' parts of the constraint surface.

	The phase space \(\tilde{\Gamma}\) is \(\mbox{\bf R}^4\) with the
natural coordinates \(q_1,q_2,p_1,p_2\) and the symplectic form is given by
\(\tilde{\Omega} = dp_1\wedge dq_1 + dp_2\wedge dq_2\). The constraint surface
is the hyperboloid given by the equation \(C=0\), where
\[ C = p_1^2 - p_2^2 - q_1^2 - q_2^2 + 1.  \]
The system is completely integrable as there are two integrals of motion that
are in involution; let us denote these integrals as follows: \(A:=(1/2)(p_1^2
- q_1^2)\) and \(B:=(1/2)(p_2^2 + q_2^2)\). Thus, the c-orbits will lie at the
cylinders \(A=\mbox{const}\), \(B=\mbox{const}\); their projections to the
\((p_1,q_1)\)-plain are hyperbolas \(A=\mbox{const}\) and those to the
\((p_2,q_2)\)-plain are circles \(B=\mbox{const}\). The general solution of
the equations of motion is easily found: the c-orbit through the point
\((\bar{q}_1,\bar{p}_1,\bar{q}_2,\bar{p}_2)\) is given by the equations:
\bea
  q_1 & = & \bar{q}_1\,\cosh t + \bar{p}_1\,\sinh t, \label{orbq1} \\
  p_1 & = & \bar{q}_1\,\sinh t + \bar{p}_1\,\cosh t, \label{orbp1} \\
  q_2 & = & \bar{q}_2\,\cos t - \bar{p}_2\,\sin t, \label{orbq2} \\
  p_2 & = & \bar{q}_2\,\sin t + \bar{p}_2\,\cos t. \label{orbp2}
\eea
The set \(E:=\{X\in\Gamma|A=0\}\) plays a very special role (at \(E\),
\(B=2\)); there is one critical c-orbit \(E_0\in E\) with \(p_1=q_1=0\)
and \(4\times S^1\) exceptional (imprisoned) orbits on the four separating
manifolds \(E_1\subset E\), \(E_2\subset E\), \(E_3\subset E\) and \(E_4\subset
E\) defined as follows:
\beann
  E_1: & p_1-q_1=0, & p_1+q_1>0, \\
  E_2: & p_1-q_1=0, & p_1+q_1<0, \\
  E_3: & p_1+q_1=0, & p_1-q_1>0, \\
  E_4: & p_1+q_1=0, & p_1-q_1<0.
\eeann
They separate the constraint surface into four quadrants \(T_{13}\),
\(T_{14}\), \(T_{23}\) and \(T_{24}\), each \(T_{ab}\) lying between the two
separating manifolds \(E_a\) and \(E_b\). \((\Gamma \setminus E_0)\) is a
(non-Hausdorf) manifold and the sets \((T_{13} \cup E_1 \cup T_{14})/\gamma\),
\((T_{23} \cup E_2 \cup T_{24})/\gamma\), \((T_{23} \cup E_3 \cup
T_{13})/\gamma\) and \((T_{24} \cup E_4 \cup T_{14})/\gamma\) form its maximal
Hausdorff submanifolds.

	The equations (\ref{orbq1}), (\ref{orbp1}), (\ref{orbq2}) and
(\ref{orbp2}) imply the following statement: Let \(\{\gamma_n\}\) be a
sequence of c-orbits within the quadrant \(T_{ab}\) that converges pointwise
to a c-orbit at \(E_a\), and let \(p\) be any point of
\(E_0 \cup E_b\), a set at the boundary of \(T_{ab}\). Then there is a
sequence \(p_n\) such that \(p_n\in\gamma_n\ \forall n\) and
\(\lim_{n=\infty}p_n = p\).
It follows that the space \(\Gamma/\gamma\) is non-Hausdorff, each two
c-orbits at \(E\) being non-separable (that is: each neighbourhood of the
first c-orbit intersects each neighbourhood of the second one, cf.\ section
\ref{sec:quant}). Moreover, it follows that each analytical perennial \(o\)
must have the form \(o=f(A,B)\). Indeed,
any {\em continuous} perennial must be constant along the set \(E\).
Consider a point \(X\in E\setminus E_0\). In a neighbourhood \(U\) of
\(X\), \(A\) and \(B\) are two independent analytical functions that are
constant along \(E\cap U\); any two other functions \(x^1\) and \(x^2\) that
form an analytical chart together with \(A\) and \(B\) in \(U\) must not be
constant along \(E\cap U\). Any analytical function \(F\) can be written in
\(U\) as \(f(A,B,x^1,x^2)\), where \(f\) is analytical. However, \(F\) will be
constant along \(E\cap U\) only if \(f\) does not depend on \(x^1\) and
\(x^2\), which proves the claim. The next consequence is that there is no
complete system of perennials (i.\ e.\ that separates separable c-orbits) that
will all be analytical: indeed, \(A\) and \(B\) are not independent on
\(\Gamma\), and we need at least two perennials to form a complete system. We
will use singular perennials and symmetries that will be associated with
transversal surfaces.

\subsection{Transversal surfaces}
\label{sec:trans}
{}From the fact that the set \(\Gamma/\gamma\) is non-Hausdorff, it follows
that
there is no {\em global} transversal surface (see \cite{timefunction}).
The next interesting kind of transversal surface is the inextensible
connected one: such surfaces play, for example, a key role in the Hawking
effect \cite{H+I}. In our model, the following four surfaces, \(\Gamma_i,\
i=1,2,3,4\), are of this kind;
their domains together cover \(\Gamma\setminus E_0\), they are all of the
topology \(\mbox{\bf R}^2\) and they can be defined by equations:
\(\Gamma_1\) by \(p_1 - q_1 = 1\), \(\Gamma_2\) by \(p_1 - q_1 = -1\),
\(\Gamma_3\) by \(p_1 + q_1 = 1\), and \(\Gamma_4\) by \(p_1 + q_1 = -1\),
together with the constraint equation, \(C=0\). Observe that the sets \({\cal
D}(\Gamma_i)\), \(i = 1,2,3,4\) coincide with the maximal Hausdorff
submanifolds. The disconnected transversal
surfaces \(\Gamma_1\cup\Gamma_2\) and \(\Gamma_3\cup\Gamma_4\) are almost
global.

	We will construct a time evolution between the surfaces
\(\Gamma_1\) and \(\Gamma_4\) and so illustrate the procedure described in
section \ref{sec:nonglob}. As for the choice of the two surfaces, let us just
remark that \(\Gamma_4\) lies in the future of
\(\Gamma_1\), if one takes seriously the time-orientation of  the c-orbits that
is defined by the Hamiltonian vector field of the constraint function \(C\).
Let the natural coordinates be \((x_1,y_1)\) on \(\Gamma_1\) and \((x_4,y_4)\)
on \(\Gamma_4\), and let the injection maps be given for \(\Gamma_1\) by
\[
   q_1 = B_1 - 1,\quad p_1 = B_1, \quad q_2 = x_1,\quad p_2 = y_1,
\]
and for \(\Gamma_4\) by
\[
   q_1 = B_4 - 1, \quad p_1 = -B_4, \quad q_2 = x_4,\quad p_2 = y_4,
\]
where \(B_i := \frac{1}{2}(y_i^2 + x_i^2),\ i=1,4\). For the pull-back
\(\Omega_i\) of the symplectic form \(\tilde{\Omega}\) we obtain simply
\(\Omega_i = dy_i\wedge dx_i\).

	The time shift \(\vartheta : \Gamma_1 \mapsto
\Gamma_4\) can be defined by
\bea
	x_4(\vartheta (x_1,y_1)) & = & x_1, \label{varthx} \\
	y_4(\vartheta (x_1,y_1)) & = & y_1. \label{varthy}
\eea
This is a ``natural'' choice, because
\(\vartheta\) is to define the same measurements at the different time levels
\(\Gamma_4\) and \(\Gamma_1\), and the coordinates \((x_i,y_i)\) coincide at
each of these surfaces
with the values of the phase functions \((q_2,p_2)\); one usually assumes that
the same symbol is used in the canonical formalism to denote a quantity which
is always measured in the same way. The dynamical map \(\rho : \Gamma_1
\mapsto \Gamma_4\) has the domain \(\Gamma_1'':=\{(x_1,y_1)\in\Gamma_1|B_1 <
\frac{1}{2}\}\) and the range \(\Gamma_4':=\{(x_4,y_4)\in\Gamma_4|B_4 <
\frac{1}{2}\}\).
\(\Gamma_1''\) coincides with \(\Gamma_1':=\vartheta^{-1}\Gamma_4'\) in this
case. Let us observe that the symplectic manifold \((\Gamma_k',\Omega_k)\)---a
disc of a finite symplectic volume---does
not admit any Lie (that is, finite-dimensional) group of symplectic
diffeomorphisms; this can be shown by studying the candidate Lie
algebras. The coordinate expression for
the map \(\rho\) can easily be obtained from the Eqs.
(\ref{orbq1}--\ref{orbp2}):
\bea
  x_4(\rho (x_1,y_1)) & = & x_1\,\cos T_1 + y_1\,\sin T_1, \label{xrho} \\
  y_4(\rho (x_1,y_1)) & = & -x_1\,\sin T_1 + y_1,\, \cos T_1 \label{yrho}
\eea
where
\be
	T_1 := \log (1 - 2B_1).
\label{T}
\ee
Thus, \(T_1\) diverges at the boundary of the domain \(\Gamma_1'\) of
\(\rho\). Finally, the map \(\chi : \Gamma_1' \mapsto \Gamma_1'\), defined by
\(\chi = \vartheta^{-1}\circ\rho\) has the following expression in the
coordinates
\bea
  x_1(\chi (x_1,y_1)) & = & x_1\,\cos T_1 + y_1\,\sin T_1, \label{xchi} \\
  y_1(\chi (x_1,y_1)) & = & -x_1\,\sin T_1 + y_1\,\cos T_1. \label{ychi}
\eea

	Next, we prove a property of the map \(\chi\) that will be important
for the quantum implementation of this map by one of the methods described in
section \ref{sec:nonglob}. Let \(f\)
be a function with a complete Hamiltonian vector field \(\xi_f\) and let
the flow of \(\xi_f\) be denoted by \(\Phi[f]_t\). Then the map \(\chi\)
satisfies the equation
\be
	\chi = \Phi[h]_1,
\label{chi}
\ee
where
\be
	h = \frac{1 - 2B_1}{2}\log\frac{1 - 2B_1}{e},
\label{h}
\ee
and \(e\) is the basis of natural logarithms. To show this property, we
consider the family of curves defined by
\bea
  x'_1 & = & \bar{x}_1\,\cos \bar{T}_1t + \bar{y}_1\,\sin \bar{T}_1t,
\label{xchit} \\
  y'_1 & = & -\bar{x}_1\,\sin \bar{T}_1t + \bar{y}_1\,\cos \bar{T}_1t,
\label{ychit}
\eea
\(t\in\mbox{\bf R}\), each starting for \(t=0\) at the point
\(\bar{x}_1,\bar{y}_1\); \(\bar{T}_1\) is the function defined by Eq. (\ref{T})
with the arguments \(\bar{x}_1\) and \(\bar{y}_1\). The tangent vector
\((\dot{x}_1,\dot{y}_1)\) to the curve at the point \((x_1,y_1)\) is
\beann
  \dot{x}_1 & = & -\bar{x}_1\bar{T}_1\sin \bar{T}_1t + \bar{y}_1\bar{T}_1\cos
	\bar{T}_1t, \\
  \dot{y}_1 & = & -\bar{x}_1\bar{T}_1\cos \bar{T}_1t - \bar{y}_1\bar{T}_1\sin
	\bar{T}_1t.
\eeann
The Eqs. (\ref{xchit}) and (\ref{ychit}) imply that \(B_1(x_1',y_1')
= B_1(\bar{x}_1,\bar{y}_1)\), so \(T_1(x_1',y_1') = T_1(\bar{x}_1,\bar{y}_1)\),
and so we obtain that
\bea
	\dot{x}_1 & = & T_1y_1, \label{dotx} \\
	\dot{y}_1 & = & -T_1x_1. \label{doty}
\eea
It follows that the curves (\ref{xchit}) and (\ref{ychit}) are identical with
the flow of
the vector field (\ref{dotx}) and (\ref{doty}); moreover, \(\chi\) is an
element of this flow. Thus, we have to find a function \(h\) such that
\beann
	\dot{x}_1 & = & \{x_1,h\}_1 = \frac{\partial h}{\partial y_1}, \\
	\dot{y}_1 & = & \{y_1,h\}_1 = -\frac{\partial h}{\partial x_1},
\eeann
where \(\{\cdot ,\cdot\}_1\) denotes the Poisson bracket of the symplectic
manifold \((\Gamma_1,\Omega_1)\) (cf section \ref{sec:singular}). An obvious
ansatz \(h=h(B_1)\) leads to the desired result, Eq. (\ref{h}).

	We have all classical maps that we need for the construction
of the time evolution between the two surfaces \(\Gamma_1\) and \(\Gamma_4\).
What is still missing are algebras of elementary perennials and/or
first-class canonical groups. We will construct some such algebras first, and
then look which groups they generate. The simplest procedure is to define the
singular perennials \(X_i\) and \(Y_i\) by their initial data along the
transversal surfaces \(\Gamma_i\) as follows:
\bea
	X_i|_{\Gamma_i} & = & x_i, \label{Xdef} \\
	Y_i|_{\Gamma_i} & = & y_i. \label{Ydef}
\eea
An easy calculation using the Eqs. (\ref{orbq1}--\ref{orbp2}) gives the
following results
\bea
  X_1 & = & q_2\,\cos T_- - p_2\,\sin T_-, \label{X1} \\
  Y_1 & = & q_2\,\sin T_- + p_2\,\cos T_-, \label{Y1}
\eea
for \(p_1 - q_1 > 0\) and
\be
	X_1 = Y_1 = 0 \label{zero1}
\ee
for \(p_1 - q_1 < 0\);
\bea
  X_4 & = & q_2\,\cos T_+ + p_2\,\sin T_+, \label{X4} \\
  Y_4 & = & -q_2\,\sin T_+ + p_2\,\cos T_+, \label{Y4}
\eea
for \(p_1 + q_1 < 0\) and
\be
	X_4 = Y_4 = 0 \label{zero4}
\ee
for \(p_1 + q_1 > 0\);
here,
\be
	T_{\pm} = \log|p_1 \pm q_1|.
\label{Tpm}
\ee
The perennials \(X_i\) and \(Y_i\) are pertinent (see section
\ref{sec:nonglob}) to the surface \(\Gamma_i\),
\(i=1,4\), and they are singular at \(p_1 - q_1 =0\) for \(i=1\) and at \(p_1
+ q_1 = 0\) for \(i=4\). Indeed, the Hamiltonian vector fields of these
perennials are complete (this is the only property of pertinent perennials
which is non-trivial to prove); we can show this as follows. The Eqs.
(\ref{X1}) and (\ref{Y1}) imply immediately that \(\{X_1,p_1-q_1\} =
\{Y_1,p_1-q_1\} = 0\). Hence, the Hamiltonian vector fields of these functions
are tangent to the planes \(p_1-q_1 = \mbox{const}\) and their integral curves
can never meet the singularity at \(p_1-q_1 = 0\). Inside these planes, the
vector fields can easily be integrated and found to be complete. The common
domain of the perennials \(X_1\) and \(Y_1\) is \(\tilde{\Gamma}\setminus
(E_3\cup E_0\cup E_4)\); together with the perennial \(B\), they generate the
four-dimensional ``harmonic oscillator Lie algebra'', which we
will denote by \(S_1\). From Eqs. (\ref{X1}) and (\ref{Y1}), a relation
follows, namely \(B=\frac{1}{2}(X_1^2+Y_1^2)\). \(S_1\) generates, in turn, the
four-dimensional harmonic oscillator group  with the same common invariant
domain. We will call this group \(G_1\). Similarly for the other two
perennials \(X_4\) and \(Y_4\): they define another copy of the harmonic
oscillator algebra \(S_4\) with the common domain \(\tilde{\Gamma}\setminus
(E_1\cup E_0\cup E_2)\) and another copy of the harmonic oscillator group
\(G_4\). Observe that the groups must be kept segregated, because the elements
of one move the domain of the other so that all transformations that result
from composition of the elements of the two groups would have no common domain
at all. The groups \(G_1\) and \(G_4\) have a common subgroup that is
generated by \(B\); in accordance with the rules of section \ref{sec:nonglob},
this subgroup can be denoted by \(G_1'\) or \(G_4'\), because it is the
subgroup that leaves the submanifolds \(\Gamma_1'\) and \(\Gamma_4'\)
invariant (without acting transitively on them).

	The definitions above imply that \(\theta S_1 = S_4\) and \(\vartheta
G_1\vartheta^{-1} = G_4\). It is easy to construct the (classical) Schroedinger
and the Heisenberg phase spaces and the time evolution according to the
prescription given in the section \ref{sec:nonglob}. We pass directly to the
quantum mechanics.

\subsection{Quantum mechanics}
As quantum mechanical counterparts of the phase spaces
\((\Gamma_k,\Omega_k)\), let us consider two Hilbert spaces \(K_k\) together
with harmonic oscillator annihilation operators \(a_k\), \(k=1,4\), acting
in the well-known way. In particular, there is a basis \(\{\psi_n^k\}\)
in \(K_k\) \(n=0,1,\ldots ,\ k=1,4\) such that
\beann
	a_k\,\psi_n^k & = & \sqrt{n\hbar}\,\psi_{n-1}^k, \\
	a_k^{\dagger}\,\psi_n^k & = & \sqrt{(n+1)\hbar}\,\psi_{n+1}^k,
\eeann
and the algebra \(S_k\) is represented on \(K_k\) by
\beann
	\hat{X}_k & = & \frac{i}{\sqrt{2}}(a_k-a_k^{\dagger}), \\
	\hat{Y}_k & = & \frac{1}{\sqrt{2}}(a_k+a_k^{\dagger}), \\
	\hat{B} & = & a_k^{\dagger}a_k + \frac{1}{2}\hbar.
\eeann
The two representations of the corresponding groups are irreducible and
equivalent; the map \(U(\vartheta )\) which realizes the equivalence and
implements the time shift \(\vartheta\) is given by
\[	U(\vartheta )\psi_n^1 = \psi_n^4. \]

	The next step is to find the subspaces \(K_k'\) which are the quantum
counterparts of the submanifolds \(\Gamma_k'\). This is straightforward: the
states \(\psi_n^k\) are the eigenstates of the operator \(\hat{B}\) with
eigenvalues \(\hbar (n+\frac{1}{2})\), and they also define the invariant
subspaces of the group  \(G_k'\). This is analogous to the classical
observable \(B\) generating the group \(G_k'\) that leaves the submanifolds
\(\Gamma_k'\) invariant so that the orbits of the group are defined by
\(B=\mbox{const}\) with \(B<\frac{1}{2}\). Thus we can identify \(K_k'\) with
the subspace spanned by the states \(\psi_n^k\) with \(\hbar (2n+1)<1\); let
us denote the projection operator onto
these subspaces by \(P_k'\). Then, we can directly implement the map \(\chi\)
because of the relations (\ref{chi}) and (\ref{h}): let us set \( U(\chi
) = \exp (i\hat{h}) \) on \(K_1'\), that is:
\be
	U(\chi )\psi_n^1 =\left(\frac{\hbar
	(2n+1)-1}{e}\right)^{-i\frac{\hbar (2n+1)-1}{2}}\psi_n^1 \label{phases}
\ee
for all \(n<(\frac{1}{2\hbar}-\frac{1}{2})\). The Schroedinger dynamics is
then defined by the evolution operator \(U(\chi )P_1'\) on \(K_1\). The
perennials that are pertinent to \(\Gamma_1'\) form just a one-dimesional
space spanned by \(B\). The time evolution of the operator \(\hat{B}\) by
\(U(\chi )\) is trivial: \(U(\chi )\hat{B}U^{-1}(\chi ) =
\hat{B}\). This is in fact all to be said about the Heisenberg picture.
However, the change of phases defined by Eq. (\ref{phases}) is  measurable:
one has to screen the ``chaos'' in the states by the operator \(P_2'\) and
then just perform measurements corresponding to the observables from the
algebra \(S_2\).

	Thus, our model nicely illustrates sections \ref{sec:singular} and
\ref{sec:nonglob}; it is intriguing, how the necessarily bizarre properties
resulted from the extremely simple definition equations of the system.

\subsection*{Acknowledgements}
Stimulating and useful discussions with A.~Ashtekar, C.~J.~Isham, J.~Kijowski,
K.~V.~Kucha\v{r}, D.~Marolf, J.~Tolar and L.~Ziewer are acknowledged.
The first version of this paper was a talk given at the conference ``New
Trends in Geometrical and Topological Methods'' in memory of W.~K.~Clifford at
Funchal, Madeira, July 30 - August 5, 1995; the author wishes to thank to the
organizers of the conference for their hospitality and to the sponsors
for the financial support. The author also acknowledges financial support from
the European Network {\em Physical and Mathematical Aspects of Fundamental
Interactions} during the stay with the Theoretical Physics Group at Imperial
College, London, where a part of this paper has been worked out.

\end{document}